\documentclass[12pt,preprint]{aastex}

\shorttitle{Irregular Mass Transfer in Low-State Polars}
\shortauthors{Pandel \& C\'ordova}

\newcommand{\xmm}{{\it XMM-Newton}}

\newcommand{\rosat}{{\it ROSAT}}

\newcommand{\xspec}{{XSPEC}}
\newcommand{\mekal}{{MEKAL}}

\newcommand{\dpm}[3]{${#1}^{+{#2}}_{-{#3}}$}

\begin{document}


\title{Irregular Mass Transfer in the Polars VV~Puppis and V393~Pavonis during the Low State}

\author{Dirk Pandel}
\affil{Department of Physics, University of California, Santa Barbara, CA 93106, USA}
\email{dpandel@xmmom.physics.ucsb.edu}
\and
\author{France A. C\'ordova}
\affil{Institute of Geophysics and Planetary Physics, Department of Physics,\\
University of California, Riverside, CA 92521, USA}


\begin{abstract}

The polars VV~Pup and V393~Pav were observed with \xmm\ during states of low accretion rate
with peak X-ray luminosities of $\sim$$1\times10^{30}$ and $\sim$$1\times10^{31}$~erg~s$^{-1}$,
respectively.
In both polars, accretion onto the white dwarf was extremely irregular, and the accretion rate varied
by more than 1 order of magnitude on timescales of $\sim$1~hr.
Our observations suggest that this type of irregular accretion is a common phenomenon in polars
during the low state.
The likely cause of the accretion rate fluctuations are coronal mass ejections or solar flares on the
companion star that intermittently increase the mass transfer into the accretion stream.
Our findings demonstrate that the companion stars in cataclysmic variables possess highly active atmospheres.

\end{abstract}



\section{INTRODUCTION}
\label{introduction}

AM Her binaries or polars are cataclysmic variables in which the strong magnetic field of the white dwarf primary
prevents the formation of an accretion disk \citep[e.g.][]{1995cvs..book.....W}.
The accretion stream from the Roche lobe-filling companion star follows the magnetic field lines
and impacts the white dwarf near a magnetic pole.
There the accreting gas forms a standoff shock and is heated to temperatures in excess of $10^8$~K.
As the shock-heated gas in the accretion column settles onto the white dwarf, it cools via the emission of
cyclotron radiation (infrared to ultraviolet) and bremsstrahlung (X-rays).
In the absence of an accretion disk, matter is transported directly from the companion star to the white dwarf,
and the observed X-ray luminosity is a direct measure of the rate at which the companion star transfers mass
into the accretion stream.

Many polars have been observed to enter low states during which little or no emission from the accretion column
is seen and which can last from several days to years.
While the cause of these low states is not known, they must be connected to a reduction of the mass transfer rate
from the companion star.
In the past, studying the weak X-ray emission from low-state polars has been difficult, but the high
sensitivity of the new generation of X-ray telescopes promises new insights into the underlying physics.
A first \xmm\ observation of a polar during the low state revealed strong flaring at X-ray and ultraviolet (UV)
energies in UZ~For \citep{2002MNRAS.336.1049P}.
In this paper, we present \xmm\ data for the two polars VV~Pup and V393~Pav, which were observed during their low states.

VV~Pup is the third cataclysmic variable that was identified as an AM~Her-type binary \citep{1977IAUC.3054....1T}.
It has been extensively studied at infrared, optical, UV, and X-ray wavelengths.
VV~Pup has a short orbital period of 100.43546~min \citep{1980ApJ...240..871S}.
It is optically faint, with $V$ magnitudes in the range 14.5--18 \citep{2001PASP..113..764D}, and
is located at a distance of $\sim$144~pc \citep{1981MNRAS.197...31B}.
The primary and secondary poles have magnetic field strengths of 31 and 54~MG, respectively
\citep{1997AN....318..111S}.
V393~Pav (RX~J1957.1--5738), which was discovered in the \rosat\ All Sky Survey,
has an orbital period of 98.8194~min, a distance of $\sim$350~pc, and a magnetic field strength of 16~MG
at the primary pole \citep{1996A&A...313..833T}.
In both polars, the main accretion region is obscured by the white dwarf during roughly half of the orbital cycle,
thus causing the X-ray flux to be strongly modulated at the orbital period.


\section{OBSERVATIONS AND DATA REDUCTION}

VV~Pup was observed with \xmm\ \citep{2001A&A...365L...1J} on 2002 November 11.
We obtained 18.0~ks of continuous X-ray data from the two EPIC MOS cameras \citep{2001A&A...365L..27T}
and 20.4~ks from the EPIC PN camera \citep{2001A&A...365L..18S}.
The three instruments were operated with the thick blocking filters and the CCDs in small window mode.
The Optical Monitor \citep{2001A&A...365L..36M} performed six exposures with a
total duration of 23.2~ks using the UVW1 filter (240--340~nm).
The Optical Monitor was operated in fast mode with a sample time of 0.5~s.

V393~Pav was observed on 2003 October 19.
We obtained 23.4~ks of X-ray data from the EPIC MOS cameras and 23.2~ks from the EPIC PN camera.
The two EPIC MOS cameras were operated in large window mode and the EPIC PN camera in small window mode.
Thin blocking filters were used for all three instruments.
The Optical Monitor performed five exposures in fast mode with a total duration of 18.2~ks using the UVW1 filter.
Because of the low signal-to-noise ratio, we did not use the data from the Reflection Grating Spectrometer
\citep[RGS;][]{2001A&A...365L...7D} for either of the two observations.

We extracted source photons from the EPIC MOS and PN data using a circular aperture with a radius of $15^{\prime\prime}$.
The count rates given in this paper have been corrected for the 65\% enclosed energy fraction of this aperture.
We included in our analysis good photon events ($\mathrm{FLAG}=0$) in the energy range 0.2--12~keV with patterns
0--12 for EPIC MOS and 0--4 for EPIC PN.
To create X-ray light curves, we applied a barycentric correction to the photon arrival times
and summed the count rates from all three EPIC instruments.
Background rates were estimated from larger regions on the same CCD as the source image.
During the VV~Pup observation, background contributed $\sim$0.01~s$^{-1}$ to the count rate in the source aperture,
except during the last 5000~s, when the background rate increased considerably and reached up to 0.15~s$^{-1}$.
During the V393~Pav observation, the background rate was fluctuating between 0.01 and 0.05~s$^{-1}$.
We extracted source photons from the Optical Monitor data using a circular aperture of $5^{\prime\prime}$ radius,
which encloses 81\% of the source photons in the UVW1 filter.
The UV background rate was $\sim$0.5~s$^{-1}$ during both observations, which amounts to $\sim$25\% for VV~Pup
and $\sim$70\% for V393~Pav.


\section{X-RAY AND UV LIGHT CURVES}

\begin{figure*}
\epsscale{1.0}
\plotone{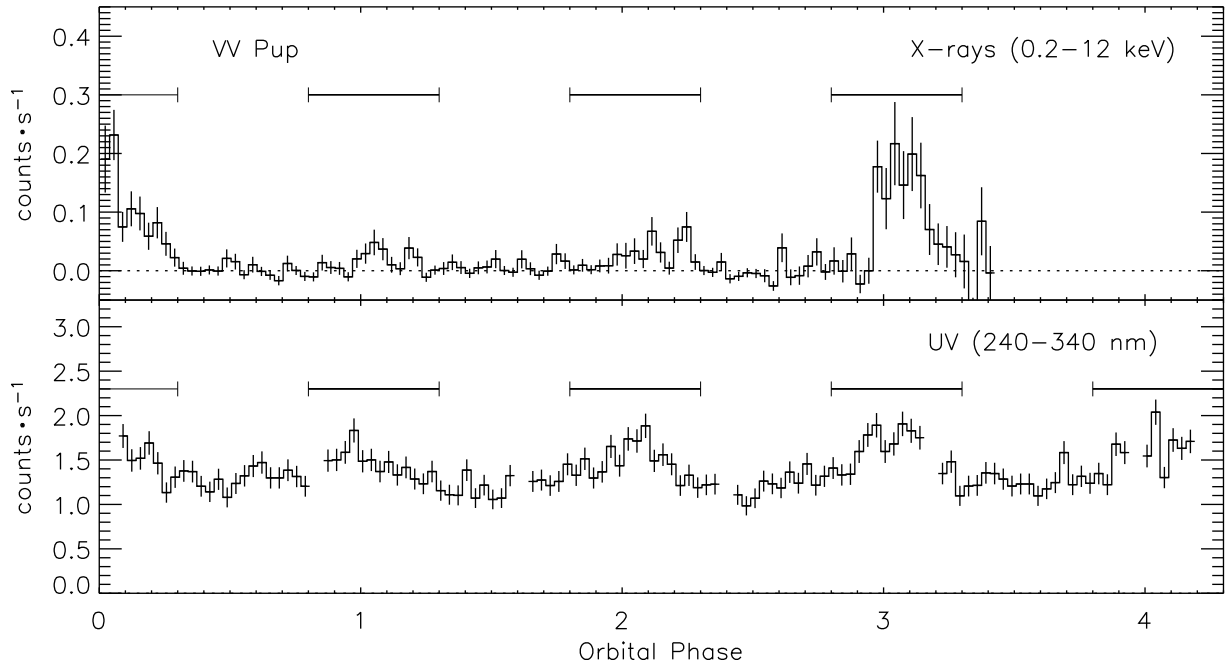}
\caption{\label{vvpuplight}
X-ray and UV light curves of VV~Pup (orbital period 100.43546~min).
Orbital phase zero is defined as BJD(TT)~2452619.780, the time of maximum optical light
as predicted by the ephemeris in \citet{1965CoKon..57....1W}.
The horizontal bars indicate the time periods during which the main accretion
region is visible.
}
\end{figure*}

\begin{figure*}
\epsscale{1.0}
\plotone{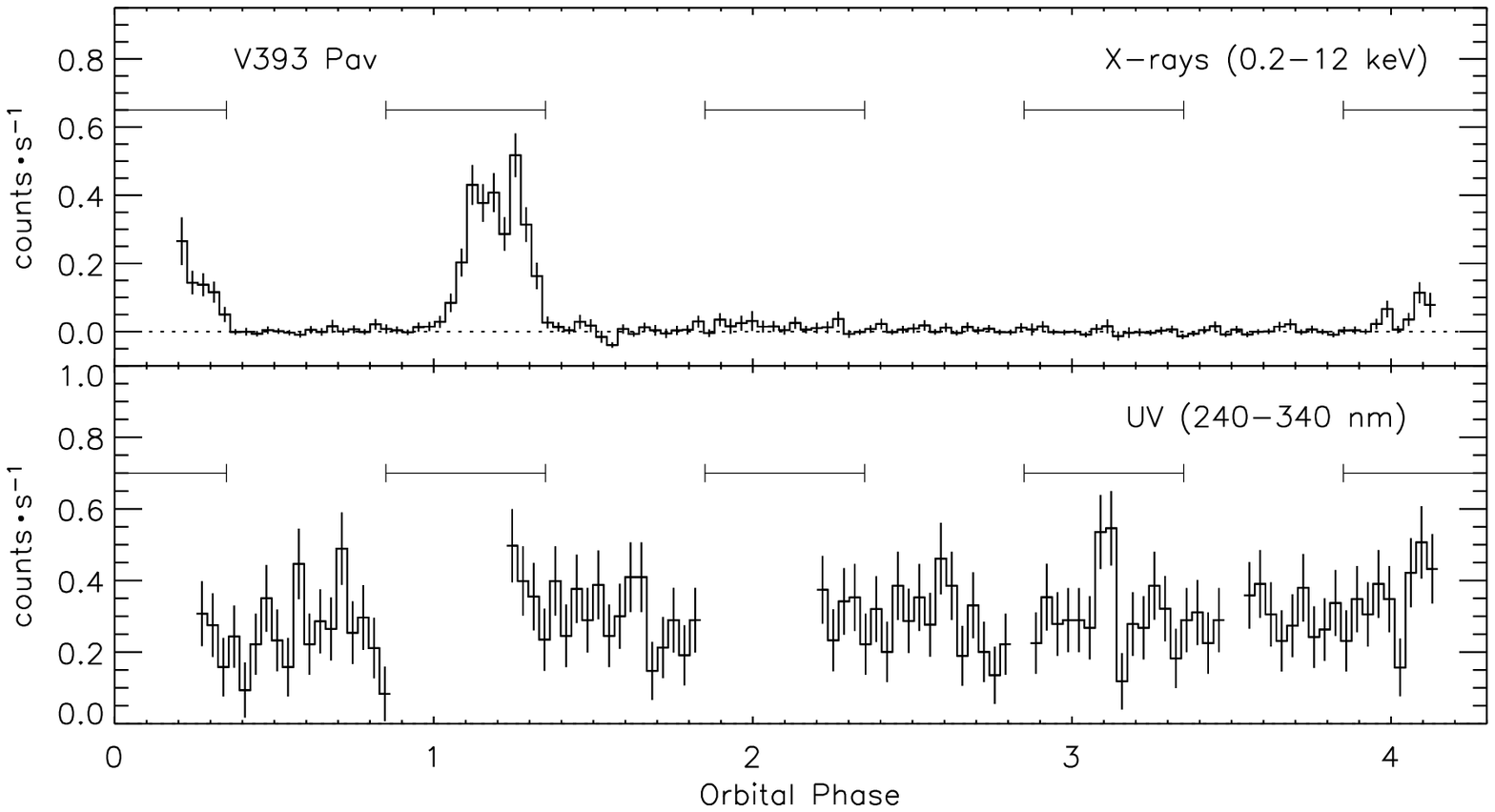}
\caption{\label{v393plight}
X-ray and UV light curves of V393~Pav (orbital period 98.8194~min).
Orbital phase zero is defined as BJD(TT)~2452931.932, the time of inferior conjunction of the secondary
as predicted by the ephemeris in \citet{1996A&A...313..833T}.
The horizontal bars indicate the time periods during which the main accretion
region is visible.
}
\end{figure*}

X-ray and UV light curves obtained from the \xmm\ data are shown in Figs~\ref{vvpuplight}
and~\ref{v393plight}.
The X-ray light curves show the sum of the count rates in the three EPIC cameras.
In the UV light curves, a count rate of 1~s$^{-1}$ corresponds to a flux of 0.13~mJy at 290~nm.
For VV~Pup, we defined as orbital phase zero BJD(TT)~2452619.780 (Barycentric Julian Date in Terrestrial Time),
which is the time of maximum optical light as predicted by the ephemeris in \citet{1965CoKon..57....1W}.
This ephemeris has an accumulated uncertainty of 0.1 in phase.
For V393~Pav, we defined as orbital phase zero BJD(TT)~2452931.932, the time of inferior conjunction of the secondary
as predicted by the ephemeris in \citet{1996A&A...313..833T}.
The accumulated uncertainty of this ephemeris is 0.15 in phase.

\xmm\ observed the two polars during states of low accretion with X-ray luminosities considerably
lower than previously seen.
\rosat\ observations of VV~Pup \citep{1996MNRAS.278..285R} found the polar in a very bright state 
with an X-ray flux that would have produced a peak count rate of 40~s$^{-1}$ in \xmm.
This is $\sim$200 times higher than the peak count rate we observed.
The X-ray flux detected with \rosat\ from V393~Pav \citep{1996A&A...313..833T} corresponds
to an \xmm\ peak count rate of 3~s$^{-1}$, which is a factor of 8 higher than that during our observation.

For our choice of orbital phase zero, the bright phases, i.e.\ the time periods during which the
main accretion region is visible, are expected to last from phase $-0.3$ to 0.2 in both polars
\citep{1984ApJ...279..785P,1996A&A...313..833T}.
Figs~\ref{vvpuplight} and~\ref{v393plight} show that the X-ray emission is mostly limited to
this phase range.
Note that the actual bright phases appear to end slightly later than the predicted phase~0.2.
This delay is, however,  within the uncertainty of the ephemerides.
We therefore assumed that the bright phases in VV~Pup and V393~Pav occur later than predicted
by 0.1 and 0.15, respectively.
The bright phases are then expected to last from $-0.2$ to~0.3 and $-0.15$ to~0.35, respectively
(indicated by horizontal bars in Figs~\ref{vvpuplight} and~\ref{v393plight}).

For VV~Pup, the more recent UV light curve in \citet{1995ApJ...445..921V} shows that,
consistent with earlier observations, the bright phase ends at $\sim$0.2.
This indicates that no significant phase change relative to the ephemeris in \citet{1965CoKon..57....1W}
has occurred.
Therefore, the delay of the bright phase by $\sim$0.1 during our observations is probably not
due to the uncertainty of the ephemeris.
The delay may, however, be the result of an accretion-rate-dependent change in the longitude of the
accretion region.
Such a longitude change has been observed in the polar WW~Hor by \citet{2002MNRAS.332..116P},
who found that, during an intermediate accretion state, the bright phase was delayed by 0.04
compared to the high state.
Similar to WW~Hor, the delay of the bright phase in VV~Pup may have been caused by the very low accretion rate
during our observation.

X-ray emission from both polars is only seen during the bright phases, which indicates that the
main accretion region is the source of the X-rays.
We find that the X-ray flux integrated over the faint phases, i.e.\ the time periods during which the accretion region
is not visible (0.3--0.8 for VV~Pup and 0.4--0.8 for V393~Pav),
is consistent with zero emission.
This implies an upper limit of 0.003~s$^{-1}$ on the average count rate during the faint phases
(at 90\% confidence level).
The X-ray flux observed during the bright phases is not constant but varies greatly by 1 order of
magnitude or more on timescales of $\sim$1~hr.
The probable cause of these brightness fluctuations is a highly variable accretion rate.

The UV light curves in Figs~\ref{vvpuplight} and~\ref{v393plight} show considerably less variability
than the X-ray light curves.
In particular, the large increases of the X-ray flux do not appear to be accompanied by enhanced UV emission.
The observed UV flux is most likely thermal radiation from the white dwarf.
On average, \xmm\ detected UV fluxes of 1.4~s$^{-1}$ (0.18~mJy) for VV~Pup and 0.30~s$^{-1}$ (0.039~mJy) for V393~Pav.
For an assumed white dwarf radius of $8\times10^8$~cm, the UV fluxes are consistent with blackbody radiation
from white dwarfs with temperatures of 11,000 and 11,500~K, respectively.
These temperatures are similar to those typically found for polars \citep[e.g.][]{1999PASP..111..532S}.
The UV light curve of VV~Pup shows a 30\% orbital modulation that peaks around phase 0.0,
near the center of the bright phase.
The modulation is likely due to a higher temperature of the white dwarf around the main accretion region.
No orbital modulation is seen for V393~Pav.


\section{X-RAY SPECTRA}
\label{spectra}

\begin{figure*}
\epsscale{0.6}
\plotone{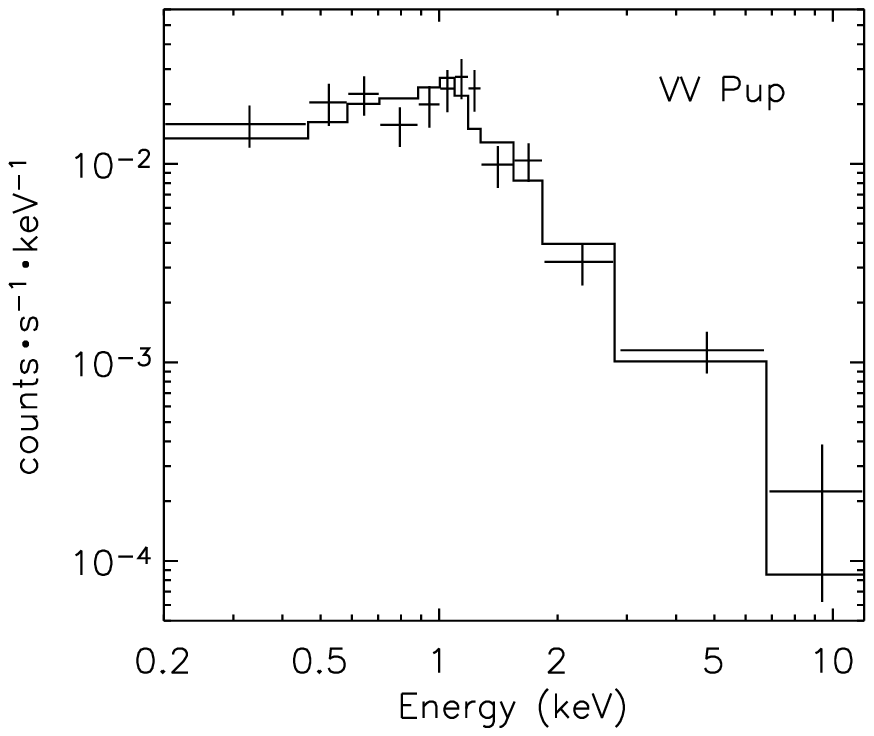}
\vspace{0.2in}
\plotone{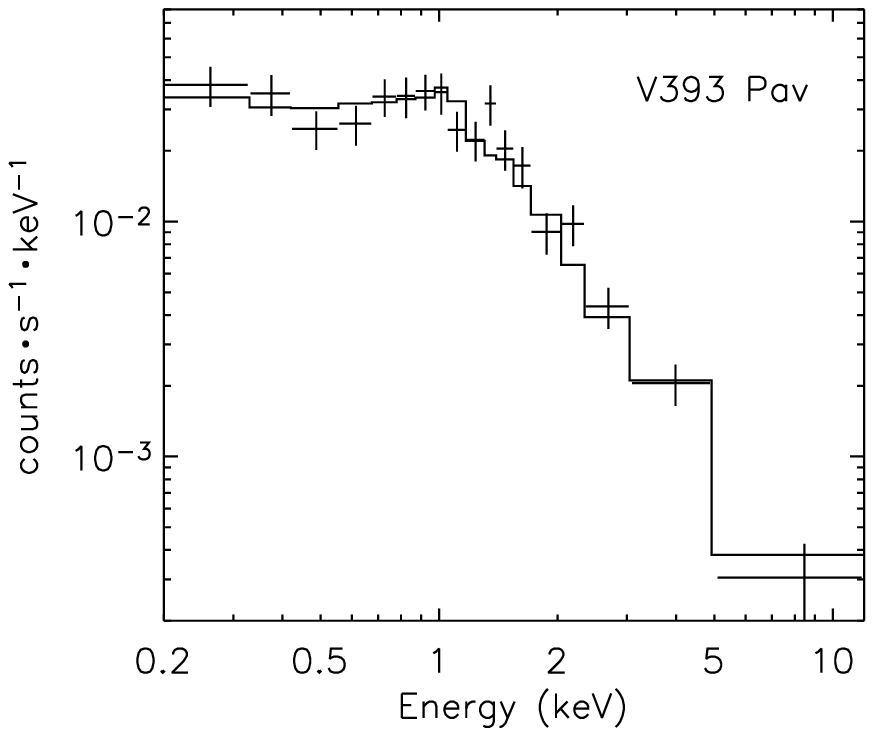}
\caption{\label{vvpupspec}
Average X-ray spectra during the bright phases.
Shown are the combined spectra of the three EPIC detectors averaged over the bright phase time intervals
indicated by the horizontal bars in Figs~\ref{vvpuplight} and~\ref{v393plight}.
The solid lines show the best fit with a cooling flow model.
}
\end{figure*}

Fig.~\ref{vvpupspec} shows the X-ray spectra of VV~Pup and V393~Pav averaged over the bright phase time intervals
(indicated by the horizontal bars in Figs~\ref{vvpuplight} and~\ref{v393plight}).
X-ray emission from polars during the high luminosity state typically consists of a strong blackbody component
at low energies and a weak bremsstrahlung component that is visible only at high energies.
In V393~Pav, the blackbody component dominated the spectrum below 0.6~keV during a previous X-ray observation
\citep{1996A&A...313..833T}.
The bremsstrahlung is thought to be emission from the shock-heated plasma in the accretion column,
while the blackbody radiation is due to reprocessing of hard X-rays in the white dwarf's atmosphere
around the accretion region.
In contrast to the high state, the spectra in Fig.~\ref{vvpupspec} show no evidence of a soft blackbody
component.
It is likely that, owing to the low accretion rates, the blackbody temperature was reduced,
and the blackbody distribution was shifted out of the detector bandpass.
The X-ray emission detected by \xmm\ appears to be solely due to the optically thin, cooling
plasma in the accretion column.

Spectral fitting was performed with the \xspec\ package, version 11.2 \citep{1996adass...5...17A}.
We binned the spectra at one-third of the FWHM detector resolution and performed
simultaneous fits to the data from all three EPIC cameras.
To account for the low number of counts per bin, we used $C$-statistic \citep{1979ApJ...228..939C}.
We fitted the spectra with three different models: a single-temperature bremsstrahlung model,
the \mekal\ model for an optically thin and collisionally ionized plasma
based on calculations by \citet{1985A&AS...62..197M} and \citet{1995ApJ...438L.115L},
and a simple cooling flow model according to \citet{1988cfcg.work...53M}.
Whereas the first two models assume a single plasma temperature $T$, the latter model describes
an isobaric cooling flow with a continuous temperature distribution.
We calculated the model spectrum of the cooling flow by adding single-temperature \mekal\ models
on a grid with spacing 0.1 in $\log T$.
Each of the three models has two free parameters.
They are, for the bremsstrahlung and \mekal\ models, the temperature $kT$ and the emission
measure $EM$ and, for the cooling flow model, the maximum or initial temperature $kT$ and
the mass accretion rate $\dot{M}$.
To calculate emission measures, accretion rates, and luminosities, we assumed isotropic emission
and neglected reflection of X-rays by the white dwarf.
As distances to the polars, we used those given in Section~\ref{introduction}.
For the \mekal\ and the cooling flow models, we fixed elemental abundances at the solar values in
\citet{1989GeCoA..53..197A}.

\begin{deluxetable}{llccccc}
\tablecaption{Spectral Fit Parameters\label{spectralfits}}
\tablewidth{0pc}
\rotate
\tablehead{
\colhead{Object} & \colhead{Model} & \colhead{$kT$}         & \colhead{$EM$}                   &
\colhead{$\dot{M}$}              & \colhead{$L_{bol}$}                & \colhead{$C$-statistic} \\
                 &                 & \colhead{(keV)}        & \colhead{($10^{52}\;$cm$^{-3}$)} &
\colhead{($10^{13}$~g~s$^{-1}$)} & \colhead{($10^{30}$~erg~s$^{-1}$)} & \colhead{(858 bins)}
}
\startdata
VV Pup    &  Bremsstrahlung  &  \dpm{2.7}{1.4}{0.8}    &  $9.7\pm1.5$   &  --            &  1.23  &  619  \\[0.6ex]
          &  \mekal          &  \dpm{3.4}{1.0}{0.8}    &  $6.5\pm0.8$   &  --            &  1.32  &  610  \\[0.6ex]
          &  Cooling Flow    &  \dpm{9.9}{6.8}{3.4}    &  --            &  $3.8\pm1.2$   &  1.46  &  612  \\[0.6ex]
\tableline\\[-1.5ex]
V393 Pav  &  Bremsstrahlung  &  \dpm{4.5}{1.8}{1.1}    &  $81\pm7\phn$  &  --            &  13.1  &  639  \\[0.6ex]
          &  \mekal          &  \dpm{4.5}{1.4}{1.0}    &  $61\pm5\phn$  &  --            &  13.8  &  639  \\[0.6ex]
          &  Cooling Flow    &  \dpm{17.1}{11.4}{5.8}  &  --            &  $25\pm7\phn$  &  16.6  &  637  
\enddata
\tablecomments{
The table shows the best-fit parameters for three spectral models (see Section~\ref{spectra}).
The parameters shown are the plasma temperature $kT$, the emission measure $EM$, the accretion rate $\dot{M}$,
the bolometric luminosity $L_{bol}$, and the likelihood according to $C$-statistic
(for assumed distances see, Section~\ref{introduction}).
Note that for the cooling flow model $kT$ is the maximum or initial temperature of the flow.
The values of $EM$, $\dot{M}$, and $L_{bol}$ have been normalized to count rates of 0.2~s$^{-1}$ for VV~Pup
and 0.4~s$^{-1}$ for V393~Pav.
}
\end{deluxetable}

The best-fit parameters for the three models are shown in Table~\ref{spectralfits}.
The likelihood values of the $C$-statistic are very close and do not favor a particular model,
although the multi-temperature cooling flow is probably the most appropriate model
for the accretion column.
$C$-statistic does not provide a measure for the quality of the fit.
However, if the spectra are rebinned as shown in Fig.~\ref{vvpupspec}, $\chi^2$-statistic can be used,
and we find good agreement between the data and the cooling flow model with
$\chi^2(dof)=12.8(11)$ for VV~Pup and $16.3(16)$ for V393~Pav.
Our estimates for the bolometric luminosity $L_{bol}$ are fairly model independent
since most of the X-ray flux is inside the 0.2--12~keV range and directly detected by \xmm.
Note that $EM$, $\dot{M}$, and $L_{bol}$ have been normalized to count rates of 0.2~s$^{-1}$ (VV~Pup)
and 0.4~s$^{-1}$ (V393~Pav), which correspond approximately to the peaks of the light curves
in Figs~\ref{vvpuplight} and~\ref{v393plight}.
Adding interstellar absorption to the spectral model did not improve the quality of the fit.
This establishes upper limits of $1\times10^{20}$~cm$^{-2}$ for VV~Pup and $2\times10^{20}$~cm$^{-2}$ for V393~Pav
on the neutral hydrogen column density $N_H$.
For VV~Pup, this is consistent with a previously found upper limit of $1\times10^{19}$~cm$^{-2}$
\citep{1996MNRAS.278..285R}.


\section{DISCUSSION}

Both polars were observed with \xmm\ during states of low accretion rate.
With a peak X-ray luminosity of $\sim$$1\times10^{30}$~erg~s$^{-1}$, VV~Pup was more than
2 orders of magnitude fainter than during a previous high-state observation
\citep{1996MNRAS.278..285R}.
For V393~Pav, we derived a peak X-ray luminosity of $\sim$$1\times10^{31}$~erg~s$^{-1}$.
Compared to the only previous X-ray observation \citep{1996A&A...313..833T},
the polar was about 1 order of magnitude fainter.
The \xmm\ spectra show no evidence of the soft blackbody component due to X-ray reprocessing
that is typically seen in polars during the high state.
Owing to the low accretion rates, this blackbody radiation was probably emitted at a lower temperature,
outside the \xmm\ bandpass.

In both polars, X-ray emission was only seen during the bright phases, while no emission was detected during
the faint phases.
This clearly shows that the X-rays originated from the main accretion region on the white dwarf.
Furthermore, the X-ray spectrum is consistent with emission from the shock-heated, optically thin plasma
in the accretion column.
As shown in Figs~\ref{vvpuplight} and~\ref{v393plight}, the X-ray luminosity of the accretion region
was fluctuating by more than 1 order of magnitude on timescales of $\sim$1~hr.
In the absence of an accretion disk, these luminosity changes must be due to a strongly varying
mass transfer rate from the companion star into the accretion stream.

Large brightness fluctuations during the low state, other than the orbital modulations, have been observed
in a few polars.
However, in only one case could these variations be unambiguously attributed to irregular mass transfer from
the companion star.
\citet{2002MNRAS.336.1049P} observed a flare with \xmm\ during an extremely low accretion state of UZ~For.
The flare, which lasted $\sim$1000~s, increased the X-ray flux by a factor of $\sim$30 and was also detected
in the UV.
The authors showed that the flare was caused by an accretion event on the white dwarf following an
intermittent increase in the mass transfer rate from the companion star.
Similar transient events have been observed by \citet{1993ApJ...414L..69W} with the {\it Extreme Ultraviolet Explorer}
in the polar QS~Tel.
As possible causes, the authors suggested a varying mass transfer rate, filaments in the accretion stream,
and solar flares on the companion star.
\citet{2000A&A...354.1003B} observed small optical flares (0.3--0.5~mag) in AM~Her during a low state.
These flares were probably caused by either accretion rate fluctuations or instabilities in the accretion stream.
Also for AM~Her, \citet{shakhovskoy1993} reported a large optical transient ($\sim$2~mag).
The properties of the transient (exponential decay, blue color, no polarization) suggest emission
from a solar flare on the companion star.
A solar flare was probably also responsible for an optical transient ($\sim$1~mag in $B$) observed
in the low accretion rate polar HS~1023+3900 by \citet{2001A&A...374..189S}.

Polars are known to exhibit strong flaring during high accretion states \citep[e.g.][]{1986MNRAS.220..633C}.
These brightness fluctuations are probably caused by dense blobs or filaments in the accretion stream
\citep{1982A&A...114L...4K,1988A&A...193..113F}.
Some of the flaring observed during high states may be due to irregular mass transfer
from the companion star.
However, it is difficult to distinguish these brightness variations from those caused by inhomogeneities
in the accretion stream.
Moreover, if the accretion rate fluctuations are of similar magnitude to those during the low state,
they would be small compared to the average high-state accretion rate and probably undetectable.

The \xmm\ observations of the three polars VV~Pup, V393~Pav, and UZ~For demonstrate that extremely irregular
mass transfer is a common phenomenon in polars during the low state.
While the causes of low states in polars are poorly understood, it is clear that they must be due to variations
of the rate at which the companion star transfers mass into the accretion stream near the L1 Lagrange point.
Variations in the size of the Roche lobe occur on timescales of $10^4$--$10^5$~yr \citep{1995ApJ...444L..37K,1988A&A...202...93R}
and cannot be responsible for the low states.
It has been suggested that starspots can pass under the L1 point and greatly reduce the mass transfer rate
\citep{1994ApJ...427..956L,1998ApJ...499..348K,2000A&A...361..952H}.
However, \citet{2000ApJ...530..904H} showed that, in short-period cataclysmic variables, the critical Roche surface
is well above the photosphere, so that the mass flow through the L1 point is actually controlled by the magnetic
activity in the chromosphere.
In both scenarios, coronal mass ejections or solar flares near the L1 point can lead to an intermittent increase
of the mass transfer rate on short timescales, thus causing the X-ray flaring observed in the three polars.
The irregular mass transfer seen during low states provides evidence
of a highly active stellar atmosphere near the L1 point.

Little is known about the level of stellar activity on the companion stars of cataclysmic variables, yet their rapid
rotation, which is synchronized with the binary orbital motion, suggests that they are highly active.
Evidence of starspots and magnetic flares on the companion stars has been found in a few cataclysmic variables
\citep[e.g.][]{2000ApJ...530..904H,2002ApJ...568L..45W,shakhovskoy1993}.
From our accretion rate measurements (Table~\ref{spectralfits}), we estimate that the large X-ray flares shown in
Figs~\ref{vvpuplight} and~\ref{v393plight} were caused by accretion of $\sim$$5\times10^{16}$~g (VV~Pup) and
$\sim$$5\times10^{17}$~g (V393~Pav) of gas onto the white dwarf.
These masses are consistent with those of solar flares on active M~dwarfs, which are in the range
$10^{15}$--$10^{18}$~g \citep[derived from density and volume measurements in][]{1990A&A...228..403P}.
Note, however, that the total accretion rates and therefore the flare masses may actually be higher than
our estimates (see next paragraph).
With the correction factors derived below, our flare mass estimates increase to $\sim$$10^{18}$~g,
which is still within the mass range of solar flares on M~dwarfs.

At the low accretion rates observed in the two polars, cooling in the accretion column is likely
dominated by cyclotron radiation \citep[e.g.][]{1979ApJ...234L.117L}.
Our measurements of $L_{bol}$ and $\dot{M}$, which are solely based on the X-ray emission,
may therefore be considerably below the total luminosities and accretion rates.
A rough estimate of the total luminosity (X-ray plus cyclotron) can be obtained from the theoretical
results in \citet{1996A&A...306..232W}.
We estimate from their fig.~9 that the total $L_{bol}$ and $\dot{M}$ are higher than those
in Table~\ref{spectralfits} by a factor of $\sim$20 for VV~Pup and $\sim$3 for V393~Pav.
Here we assumed a typical fractional area of the accretion region $f=10^{-3}$ \citep[e.g.][]{1998ApJ...506..824S},
a white dwarf mass of 0.7~$M_\odot$, and a white dwarf radius of $8\times10^8$~cm.
The correction factors imply, for both polars, a specific accretion rate of $\sim$0.1~g~cm$^{-2}$~s$^{-1}$
at the peak of the light curves.

The X-ray flare observed in UZ~For by \citet{2002MNRAS.336.1049P} was accompanied by a simultaneous UV flare
probably due to cyclotron radiation.
However, no such UV flaring is visible in the light curves of VV~Pup and V393~Pav.
Cyclotron radiation from polars is typically seen at optical and infrared wavelengths and only under
certain conditions in the near UV.
For a specific accretion rate $\dot{m}\approx0.1$~g~cm$^{-2}$~s$^{-1}$,
fig.~5 in \citet{1996A&A...306..232W} indicates that the cyclotron flux at 290~nm
($1.0\times10^{15}$~Hz) has already dropped well below the peak in the infrared,
and it is probably much weaker than the thermal radiation from the white dwarf (not shown in the figure).
We therefore do not expect to see significant cyclotron emission in the UV light curves of VV~Pup and V393~Pav.
In UZ~For, some cyclotron radiation may have been emitted in the near UV because of the 
higher magnetic field strength ($\sim$50~MG) and possibly a smaller accretion region.

As indicated by the presence of a hot, X-ray emitting plasma, 
the specific accretion rate $\dot{m}\approx0.1$~g~cm$^{-2}$~s$^{-1}$ places the two polars in the regime
of a standoff shock \citep[e.g.][]{1996A&A...306..232W}.
Recently, several polars have been discovered that appear to be in a permanent state of very low accretion
rate with $\dot{m}\approx10^{-3}$~g~cm$^{-2}$~s$^{-1}$
\citep{1999A&A...343..157R,2000A&A...358L..45R,2003ApJ...583..902S}.
At this low $\dot{m}$, a standoff shock does not form, and accretion is instead characterized by the bombardment
solution \citep{1982A&A...114L...4K,1996A&A...310..526R}.
While these low accretion rate polars are strong emitters of cyclotron radiation, their X-ray emission
is very weak.
Unfortunately, the observational data on these objects are insufficient to determine whether
they exhibit accretion rate fluctuations similar to those in VV~Pup and V393~Pav.
Variability studies of the low accretion rate polars could reveal a possible correlation between the average
mass transfer rate and the level of stellar activity on the companion star.


\section{CONCLUSIONS}

\xmm\ observations of VV~Pup and V393~Pav found the polars in states of low accretion rate with peak X-ray
luminosities of $\sim$$1\times10^{30}$ and $\sim$$1\times10^{31}$~erg~s$^{-1}$, respectively.
In both polars, accretion onto the white dwarf was highly variable and fluctuated by more than
1 order of magnitude on timescales of $\sim$1~hr.
All of the observed X-ray emission appears to originate from the main accretion region.
The X-ray spectrum is consistent with emission from the shock-heated plasma in the accretion column
but does not show evidence of the soft blackbody component that is commonly seen during high states.
The UV flux is considerably less variable than the X-ray flux, and no brightness fluctuations simultaneous
to those in the X-ray data are seen.
For VV~Pup, we found a 30\% orbital UV modulation due to the heating of the polar region
by the accretion stream.

From our observations of VV~Pup and V393~Pav and from the \xmm\ data of UZ~For presented in \citet{2002MNRAS.336.1049P},
it becomes evident that extremely irregular accretion is a common phenomenon in polars during the low state.
In the absence of an accretion disk, the cause of these accretion rate fluctuations must be a strongly varying
mass transfer rate from the companion star into the accretion stream.
The large variations of the mass transfer rate are likely a result of coronal mass ejections or solar flares
near the L1 Lagrange point.
Our findings demonstrate that the companion stars in cataclysmic variables possess highly active atmospheres.
The high sensitivity of \xmm\ provides a new way to study stellar activity on the companion stars in polars.


\acknowledgments

This work is based on observations obtained with \xmm, an ESA science mission
with instruments and contributions directly funded by ESA member states and the
USA (NASA).
The authors acknowledge support from NASA grant NAG5-12934.


\bibliographystyle{apj}
\bibliography{vvpup}

\end{document}